\documentclass[12pt]{article}
\usepackage{epsfig}
\newcommand{\nc}{\newcommand}
\nc{\postscript}[2] 
{\setlength{\epsfxsize}{#2\hsize}\centerline{\epsfbox{#1}}}
\nc{\bg}{B. Grzadkowski}
\nc{\non}{\nonumber}
\nc{\hc}{\hbox {h.c.}} \nc{\re}{\hbox {Re}} 
\nc{\mev}{\hbox {MeV}} \nc{\gev}{\;\hbox {GeV}} \nc{\tev}{\;\hbox {TeV}}
\def\lsim{\mathrel{\raise.3ex\hbox{$<$\kern-.75em\lower1ex\hbox{$\sim$}}}}
\def\gsim{\mathrel{\raise.3ex\hbox{$>$\kern-.75em\lower1ex\hbox{$\sim$}}}}

\nc{\prd}[3]{{\it Phys.\ Rev.}\ {{\bf D{#1}} (#2), #3}}
\nc{\prl}[3]{{\it Phys.\ Rev.\ Lett.}\ {{\bf {#1}} (#2), #3}}
\nc{\plb}[3]{{\it Phys.\ Lett.}\ {{\bf B{#1}} (#2), #3}}
\nc{\npb}[3]{{\it Nucl.\ Phys.}\ {{\bf B{#1}} (#2), #3}}
\nc{\ptp}[3]{{\it Prog.\ Theor.\ Phys.}\ {{\bf {#1}} (#2), #3}}
\nc{\zfp}[3]{{\it Z.\ Phys.}\ {{\bf C{#1}} (#2), #3}}
\nc{\epj}[3]{{\it Eur.\ Phys.\ J.}\ {{\bf C{#1}} (#2), #3}}
\nc{\mpla}[3]{{\it Mod.\ Phys.\ Lett.}\ {{\bf A{#1}} (#2), #3}}
\nc{\rmp}[3]{{\it Rev.\ Mod.\ Phys.}\ {{\bf {#1}} (#2), #3}}
\nc{\ijmpa}[3]{{\it Int.\ J.\ of\ Mod.\ Phys.}\
               {{\bf A{#1}} (#2), #3}}
\nc{\Lsp}{\;\;\;\;\;\;\;\;\;\;}  \nc{\LLLsp}{\lspace \lspace}
\nc{\lsp}{\;\;\;\;\;\;}
\nc{\spac}{\;\;\;}
\nc{\noi}{\noindent}
\nc{\beq}{\begin{equation}}   \nc{\eeq}{\end{equation}}
\nc{\bea}{\begin{eqnarray}}   \nc{\eea}{\end{eqnarray}}
\nc{\baa}{\begin{array}}      \nc{\eaa}{\end{array}}
\nc{\bit}{\begin{itemize}}    \nc{\eit}{\end{itemize}}
\nc{\ben}{\begin{enumerate}}  \nc{\een}{\end{enumerate}}
\nc{\bce}{\begin{center}}     \nc{\ece}{\end{center}}

\def\lam{\Lambda}

\def\lcal{{\cal L}}
\def\sq2{\sqrt{2}}
\def\arg{(m_ny)}
\def\ph{\varphi}
\def\ms{m^2(\ph)}
\def\m4{m^4(\ph)}
\def\mn2{m_n^2}

\def\nm{n_{max}}

\begin{document}
\pagestyle{plain}
\pagestyle{empty} \setlength{\footskip}{2.0cm}
\setlength{\oddsidemargin}{0.5cm} \setlength{\evensidemargin}{0.5cm}
\renewcommand{\thepage}{-- \arabic{page} --}
\def\mib#1{\mbox{\boldmath $#1$}}
\def\bra#1{\langle #1 |}      \def\ket#1{|#1\rangle}
\def\vev#1{\langle #1\rangle} \def\dps{\displaystyle}
\nc{\tb}{\stackrel{{\scriptscriptstyle (-)}}{t}}
\nc{\bb}{\stackrel{{\scriptscriptstyle (-)}}{b}}
\nc{\fb}{\stackrel{{\scriptscriptstyle (-)}}{f}}
\nc{\pp}{\gamma \gamma}
\nc{\pptt}{\pp \to \ttbar}
\nc{\barh}{\overline{h}}
   \def\thebibliography#1{\centerline{REFERENCES}
     \list{[\arabic{enumi}]}{\settowidth\labelwidth{[#1]}\leftmargin
     \labelwidth\advance\leftmargin\labelsep\usecounter{enumi}}
     \def\newblock{\hskip .11em plus .33em minus -.07em}\sloppy
     \clubpenalty4000\widowpenalty4000\sfcode`\.=1000\relax}\let
     \endthebibliography=\endlist
   \def\sec#1{\addtocounter{section}{1}\section*{\hspace*{-0.72cm}
     \normalsize\bf\arabic{section}.$\;$#1}\vspace*{-0.3cm}}
\vspace*{-2cm}
\begin{flushright}
$\vcenter{
\hbox{IFT-03-09}
\hbox{April, 2003}
}$
\end{flushright}
\vskip 2cm
\begin{center}
{\large\bf The Effective Potential and Vacuum Stability \\ within Universal Extra Dimensions}
\end{center}

\vspace*{1cm}
\begin{center}
\renewcommand{\thefootnote} {\alph{footnote})}
{\sc Patrizia BUCCI}\footnote{E-mail address:
\tt patrizia.bucci@fuw.edu.pl} and 
{\sc Bohdan GRZADKOWSKI}\footnote{E-mail address:
\tt bohdan.grzadkowski@fuw.edu.pl}\\
Institute of Theoretical Physics, Warsaw 
University,\\
 Ho\.za 69, PL-00-681 Warsaw, POLAND\\
\vskip .6cm
\end{center}

\vskip 1.5cm

\centerline{ABSTRACT} 
\vskip .5cm

The one-loop effective potential calculated for a generic
model that originates from 5-dimensional theory
reduced down to 4 dimensions is considered. The cut-off and dimensional regularization
schemes are discussed and compared. It is demonstrated that the prescriptions are
consistent with each other and lead to the same physical consequences. Stability of the ground state is 
discussed for a U(1) model that is supposed to mimic the Standard Model extended to
5 dimensions. 
It has been shown that fermionic Kaluza-Klein modes can dramatically influence the shape of the
effective potential shifting the instability scale even by several orders of magnitude.

\vfill



\newpage
\renewcommand{\thefootnote}{\arabic{footnote}}
\pagestyle{plain} \setcounter{footnote}{0}
\baselineskip=21.0pt plus 0.2pt minus 0.1pt
\section{Introduction}

For some time there has been increased interest in possible extensions of the Standard Model (SM)
that allow for fields living in extra dimensions. One possible scenario, referred to as the 
Universal Extra Dimensions (UED) model \cite{Appelquist:2000nn} assumes that all 
the SM degrees of freedom
propagate in compactified extra dimension of the size of $R\sim \tev^{-1}$ \footnote{The 
first studies of possible effects of SM fields living in TeV-scale extra dimensions  were 
performed by I.~Antoniadis \cite{Antoniadis:1990ew}.} . It has been shown that 
in fact $R^{-1}$ as low as $\sim 0.3 \tev$ is allowed by the precision electroweak observables
\cite{Appelquist:2000nn}. 
Constraints from flavor changing processes  
have been carefully analyzed in refs.\cite{Agashe:2001xt},\cite{Buras:2002ej} while
the anomalous magnetic moment has been studied in ref.\cite{Agashe:2001ra}.
All the analysis conclude that even $R^{-1} \sim 0.3 \tev$ is consistent with the existing
experimental data.
The main reason for the suppression of extra contributions to the above observables 
is the momentum conservation in the fifth dimension. In the equivalent
4D theory this implies that an emission of a single non-zero Kaluza-Klein (KK) mode is forbidden.
Consequently there is no tree-level  contributions to the electroweak observables, and
therefore KK effects are suppressed. However, the
large size of $R$ could lead to exciting phenomenology
at the next generation of colliders~\cite{Macesanu:2002db}.

Constraints from the precision electroweak observables on the Higgs physics have been 
analyzed in refs.~\cite{Appelquist:2000nn} and \cite{Appelquist:2002wb}. In particular the 
ref.~\cite{Appelquist:2002wb} shows the allowed region for the Higgs-boson mass $m_h$ and the
compactification radius $R$ in the 5D UED compactified on $S^1/Z_2$. It turns out
that for $m_h\sim 0.9\tev$ even $R^{-1} \sim 0.25 \tev$ is allowed. Since effects of KK modes
appear at the 1-loop therefore one could expect their relevance for processes that emerge
at the 1-loop level in the SM, an illustration of that reasoning could be found in
refs.~\cite{Agashe:2001xt},\cite{Buras:2002ej} and \cite{Agashe:2001ra}.
Here we will consider influence of extra dimensional physics on the stability of the ground state.
It is well known that within the SM model~\cite{bounds_SM}
and variety of its extensions~\cite{bounds_ext} contributions from fermionic degrees
of freedom could lead to an effective potential that is unbounded from below, provided
the Higgs boson mass is small enough~\cite{vacuum_stability}. 
That implies an lower bound on $m_h$ as a function of the
cut-off scale below which the theory is supposed to be stable. Since the compactification
of the 5D theory leads
to existence of an infinite tower of 4D fermions, therefore it is natural to expect that the SM
picture of the effective potential will be 
modified\footnote{For earlier discussion of the instability within extra dimensional 
theories see ref.\cite{Elizalde:1994cg}.}. Indeed, as we have found the influence of
fermionic KK modes on the scale of stability is dramatic, {\it the scale could be shifted by many 
orders of magnitude}!

The paper is organized as follows. In Section~\ref{gen_pot}, we discuss generic properties 
of the effective potential both in the cut-off and the dimensional regularization.
Section~\ref{u1model} presents details of the 5D model considered here and
also analytical results for the effective potential. 
In Section~\ref{results}, we discuss numerical results.
Concluding remarks are given in Section~\ref{summary}.

\section{The generic effective potential}
\label{gen_pot}

Here we will present results for a contribution to the one-loop effective potential
coming from an infinite tower of virtual KK modes (numbered by an integer $n$).
The following generic formula is applicable both for fermions and 
bosons circulating\footnote{For vector bosons the Landau gauge should be adopted, while for
fermions extra minus sign must be added.}
in loops:
\beq
V(\ph)=\frac12 \int\frac{d^4p}{(2\pi)^4}\sum_{n=-\infty}^{\infty}\ln[l^2E^2+(n+\omega)^2\pi^2]\,,
\label{generic}
\eeq
where  
$\omega$ is a constant shift, $E^2\equiv p^2+\ms$, 
$\ms$ is the background field dependent mass squared 
of virtual KK modes,
the momentum $p$ is defined in the Euclidean space 
($p^2=p_0^2+(\vec{p})^2$), the field independent factor $l\equiv \pi R$ 
was introduced for dimensional reasons
and all unnecessary constant terms have been dropped.

\subsection{Divergences}
There are two sources of possible divergences appearing in the effective potential (\ref{generic}):
{\it i}) the momentum integration, and {\it ii}) the infinite sum over KK modes.
The integral could be regularized either by the dimensional method or by the cut-off, while for
the sum  one can, for instance, use the method adopted by Delgado, Pomarol and Quir\'os (DPQ)
in ref.~\cite{Delgado:1998qr}, the $\zeta$ regularization (see e.g.~\cite{DiClemente:2001ge})
or just truncation of the series (for the discussion see 
refs.~\cite{Ghilencea:2001bv},\cite{Ghilencea:2001ug}). 

There is a comment here in order. 
Since both the integration and the summation are not convergent therefore
the interchange of their ordering seems to be a non-trivial issue. This question was already addressed
in ref.~\cite{DiClemente:2001ge} in the framework of 5D SUSY model compactified on the 
orbifold $S^1/(Z_2\times Z_2^\prime)$. The authors computed the effective potential performing first
the integration with dimensional regularization and then adopting the $\zeta$ regularization for the KK sum.
It has been shown that when dimensional regularization
is adopted\footnote{The effective potential found in  ref.~\cite{Delgado:1998qr} was ultraviolet
divergent, however note that the cut-off regularization was adopted there. It is easy to see that
for the dimensional regularization the result would be finite.}
 then both orderings lead to the same ultraviolet finite
result separately for scalars and fermions.
So, the ``KK regularization'' used both in ref.~\cite{Delgado:1998qr} and 
in ref.~\cite{DiClemente:2001ge} leads to the same result. 
However this regularization seems to suffer from certain drawbacks:
\bit
\item Since the 5D theory is non-renormalizable therefore
there must exist certain physical cut-off $\lam_5$, related to
the scale of more fundamental high-energy physics, e.g. string theory.
Therefore performing loop expansion in 5D it would 
be natural to cut all loop integrals $d^5p$ at the scale $\lam_5$. From the 4D perspective
the summation over KK modes corresponds to the integration over the fifth momentum 
component, so it seems to be appropriate
to limit the sum to $n\lsim \lam_5 R$, what would roughly guarantee that we sum all modes
that are lighter than the cut-off. In contrast to this strategy 
the KK-regularization requires summation over all the
modes, therefore its physical meaning seems to be rather unclear\footnote{An interesting observation 
has been made in refs.~\cite{Ghilencea:2001bv},\cite{Ghilencea:2001ug}, where the authors
showed that the vanishing of quadratic divergences that happens 
separately for bosons and fermions is a consequence of cancellation between contributions of
states of mass larger than the cut-off $\lam_5$ and light states laying below the cut-off.}.
\item The ref.~\cite{DiClemente:2001ge} shows that for the KK-regularization 
the resulting effective potential in the limit  $R\to 0$ is different when we decompactify ($R\to 0$)
before the regularization (assuming that all non-zero KK modes decouple in this case
one recovers the 4D effective potential 
generated just by the zero mode) and after the regularization (the KK-regularized 
effective potential diverges in this limit).
\eit
In this paper we are going to discuss vacuum stability,  
so for a given mass of the Higgs boson zero mode we will determine the scale below which
the model makes sense (the vacuum is stable). Therefore it seems to be meaningful to restrict
the mass spectrum of the KK modes to those which are lighter than the cut-off, so in the 
following we will 
also consider truncation of series over KK modes to those $n<n_{max}\equiv{\lam_5 R}$. From
the 5D perspective, this will
correspond to a cut-off for the integration over the fifth momentum component. 
Then, of course, the sum is finite and therefore question of ordering for the summation 
and integration becomes meaningless.
Concerning the regularization of the $d^4p$ integral the analogous approach would be to
adopt a cut-off regulator. We will illustrate this strategy below. 

Even though the cut-off regularization seems to be the most natural one,  
there exist also arguments against it. The standard 
objections are the following:
\bit
\item Because of the compactification on the circle, the shift along the extra direction;
$y \to y + 2\pi R$ should 
leave the theory unchanged. Therefore the fifth component of momentum is quantized 
to be elements of ${\cal Z}/R$. A consequence of that is the ``integer shift'' symmetry,
i.e. a symmetry under an integer shift of KK modes. Obviously, cutting the series 
breaks the symmetry, as there would be no modes to go.
\item Another drawback of the regularization through a limited number of modes is
the fact that  5D gauge invariance is broken in that case. Namely, limiting the number of KK 
modes we impose a condition on the 5D gauge transformation parameter $\theta(x,y)$
that has the following general expansion:
\beq
\theta(x,y)=\frac{1}{\sqrt{2\pi R}}\left[\theta_0(x)+
\sq2\sum_{n=1}^{\infty}\theta_n(x)\cos\arg\right]\,.
\eeq
Therefore, 
if we had summed up to $n_{max}$, then obviously, the series would not be able to reproduce
all possible 5D gauge parameter functions $\theta(x,y)$.
\eit
So, it is essential to look for a regularization prescription that would be consistent with
all the symmetries that are present. The dimensional regularization
is the standard option that satisfy the requirement.
An interesting and natural generalization of dimensional regularization for sums over KK modes
was developed 
in refs.~\cite{GrootNibbelink:2001bx},\cite{GrootNibbelink:2001hq}. 
The strategy is in its spirit similar to the method
adopted earlier by DPQ in ref.~\cite{Delgado:1998qr}, namely the sum
could be traded for a one-dimensional contour integral that one can regularize by 
analytic continuation 
in the number of dimensions. The great advantage of this approach is that both the gauge and also
the ``integer shift'' symmetries are preserved. 

Therefore for completeness and comparison 
we will consider in the following sections 
the effective potential found adopting both the cut-off regularization
with limited KK-summation and  the KK
regularization~\cite{Delgado:1998qr} proposed by DPQ\footnote{As it will be discussed shortly
the KK regularization leads to the same result as the dimensional regularization of the sum
over KK modes
and of the integral along the line proposed in 
refs.~\cite{GrootNibbelink:2001bx},\cite{GrootNibbelink:2001hq}.}.

\subsection{Limited KK-summation and cut-off regularization}
\label{cut-off}
In this section we will discuss an effective potential within a 5D theory of a scalar field
assuming that only a zero mode (in KK expansion) of the scalar
can acquire a vacuum expectation value: $\ph$.
Because of later applications we will restrict ourself to the sum over non-negative $n$ 
and $\omega=0$ in the effective potential  (\ref{generic}). Then for a limited number of KK modes
with the 4D cut-off ($\Lambda$) regularization the effective potential reads:
\beq
V_{eff\, bare}^{1-loop}=
\frac{1}{32\pi^2}\sum_{n=0}^{n_{\max}}\left\{\lam^2\ms+
\frac12[\ms+\mn2]^2\left[\ln\left(\frac{\ms+\mn2}{\lam^2}\right)-\frac12\right]\right\}\,,
\label{potlam}
\eeq
where $\mn2\equiv (n/R)^2$ and $n_{max}\equiv{\lam_5 R}$ for $\Lambda_5$ being the 5D cut-off
of the $dp_5$ integration. Therefore
imposing such a limit on the number of modes is roughly equivalent to 5D cut-off
regularization of $dp_5$ integration.
The terms that are divergent in the limit $\lam \to \infty$ are the following
\bea
V_{eff}^{1-loop}|_{div}&=&\frac{(n_{max}+1)}{32\pi^2}\left\{\ms[\lam^2+\right.\\
&&\left.-\frac{n_{max}}{3R^2}(1 + 2n_{max})
\ln(R\lam)]-\m4\ln(R\lam)\right\}\,.\non
\label{divcut}
\eea
There is a comment here in order. In a case of mixing between virtual degrees of freedom, 
non-diagonal mass matrices
may appear and the eigen values are in general
non-polynomial functions of $\ph$ (see for example the $(A_{5\,n},\chi_n)$-system for the model 
discussed in sec.\ref{u1model}). At first sight this seems to jeopardize 
the process of renormalization
since only $\ph^2$ and $\ph^4$ counter-terms are at our disposal while the divergent
contributions appear to be non-polynomial functions of $\ph$.
However, for a general mass matrix we should replace $\ms$ and $\m4$ that appear in 
$V_{eff}^{1-loop}|_{div}$ by $Tr[\ms]$
and $Tr[\m4]$, respectively.  Since $Tr[\cdots]$ is invariant under diagonalization, one may use
the non-diagonal basis here, then, because all  elements of the initial non-diagonal
mass matrix squared are in general quadratic in $\ph$, therefore the counter-terms 
at hand turns out to be sufficient to remove all the divergences.  

Let us now specify the theory as a  real $\phi^4$ theory in 5D defined by the following potential:
\beq
V_{tree}=\frac12 \mu^2 \phi^2+\frac14 \lambda_5 \phi^4
\label{vphi4}
\eeq
Note that $\lambda_5$ is dimension-full and $\phi$ has dimension of ${\rm mass}^{3/2}$, while
$\ph$ (the classical zero-mode scalar field) has dimension of mass.
After reducing to 4D the tree level bare potential for the classical field $\ph$ is the following:
\beq
V_{tree}=\frac12 \mu^2 \ph^2+\frac14 \lambda \ph^4
\label{v4D}
\eeq
where now $\lambda$ is dimensionless. 
In order to remove the divergent contributions
one has to adopt appropriate 
counter-terms. The renormalization conditions that we will choose are the following:
\beq
\frac{d^2V_{eff}}{d\ph^2}|_{\ph=0}=\mu_r^2, \lsp \frac{d^4V_{eff}}{d\ph^4}|_{\ph=0}=3!\,\lambda_r
\label{rencon}
\eeq
for the 4D tree-level potential shown in (\ref{vphi4}).
The bare parameters $\mu^2$ and $\lambda$ are related to the renormalized ones and to the 
counter-terms through: 
\beq
\mu^2=\mu_r^2+\delta \mu^2, \lsp \lambda=\lambda_r+\delta \lambda\,.
\eeq
In the case of the potential (\ref{vphi4}) we have the following form of $\ms$
\beq
\ms=\frac12 \frac{d^2\,\ms}{d\,\ph^2}_{|\,\ph=0}\ph^2+m^2(0)
\label{m2}
\eeq
It is straightforward to prove that the conditions (\ref{rencon}) lead to the following counter-terms:
\bea
\delta \mu^2&=&-\frac{d^2V_{eff}^{1-loop}}{d\ph^2}_{|\, \ph=0}=\\
&&-\frac{1}{32\pi^2}\sum_{n=0}^{n_{\max}}
\frac{d^2\,\ms}{d\,\ph^2}_{|\,\ph=0}\left[\lam^2+(m^2(0)+\mn2)\ln\left(\frac{m^2(0)+\mn2}{\lam^2}\right)\right]\non\\
\delta \lambda &=&-\frac{1}{3\, !}\frac{d^4V_{eff}^{1-loop}}{d\ph^4}_{|\, \ph=0}=\\
&&-\frac{1}{64\pi^2}\sum_{n=0}^{n_{\max}}\left(\frac{d^2\,\ms}{d\,\ph^2}\right)^2_{|\,\ph=0}
\left[1+\ln\left(\frac{m^2(0)+\mn2}{\lam^2}\right)\right]\,.\non
\eea
It could be easily verified that the above counter-terms do cancel  the divergences in 
$V_{eff}^{1-loop}|_{div}$, note that the form of $m^2(\ph)$ given in (\ref{m2})
is essential for the cancellation.
Eventually, the renormalized 1-loop contribution to the effective potential reads:
\bea
V_{eff\, ren\;I}^{1-loop\; \lam}&=& \frac{1}{32\pi^2}\sum_{n=0}^{n_{\max}}
\left\{\frac12\left(\ms+\mn2\right)^2\ln\left(\frac{\ms+\mn2}{m^2(0)+\mn2}\right)+\right.\non \\
&&\left.-\frac34 m^4(\ph) + m^2(\ph)\left[m^2(0)-\frac12\mn2\right]\right\}
\label{effpotren1}
\eea

As it was already mentioned for general non-diagonal mass matrices the condition (\ref{m2})
does not hold. 
Nevertheless, as we have already discussed above the renormalization procedure could be successfully
performed.
Then it would be convenient to split the counter-terms into divergent and 
finite parts. Since  the divergent contributions to the effective potential
are linear functions of $\ms$ and $\m4$ only, therefore
they can be replaced by $Tr[\ms]$ and $Tr[\m4]$, respectively and for them (in the non-diagonal basis)
the form (\ref{m2}) holds. However, for finite parts the renormalization conditions (\ref{rencon})
turns out to be very inconvenient as they lead to quite complicated expressions for the renormalized
effective potential, therefore one can modify the above renormalization prescription 
such that one will only keep the divergent parts of $\delta m^2$ and $\delta \lambda$.
However, we will not discuss this renormalization prescription hereafter.

\subsection{Dimensional regularization }

It will be useful to repeat the derivation of the effective potential 
proposed by DPQ~\cite{Delgado:1998qr} and compare with the dimensional 
regularization of the KK sum adopted in ref.~\cite{GrootNibbelink:2001bx}.
In order to find $V(\ph)$ defined in eq.(\ref{generic}) we first define
\beq
W=\frac12\sum_{n=-\infty}^{\infty}\ln\left[(lE)^2+(n+\omega)^2\pi^2\right]\,.
\eeq
Instead of $W$ we calculate 
\beq
\frac{\partial W}{\partial E}=l^2E\sum_{n=-\infty}^{\infty}\frac{1}{(lE)^2+(n+\omega)^2\pi^2}
\label{derw}
\eeq
that is already convergent. By that procedure, an infinite, but constant ($E$-independent) 
term was dropped. This is, of course, legal, since the constant is $\ph$ independent and therefore
its elimination corresponds to the renormalization of the cosmological constant.
Then replacing the infinite sum in (\ref{derw}) by an integral in the complex plane and applying the
residues theorem to perform the integral leads to the following result:
\beq
W=lE+\frac12\left\{\ln\left(1-re^{-2lE}\right)+\ln\left(1-r^{-1}e^{-2lE}\right)\right\}\,,
\label{resw}
\eeq
where $r\equiv e^{-2i\omega\pi}$.
The first term in (\ref{resw}), that is the limit of the full $W$ when $R\to \infty$, leads to the 
effective potential for the uncompactified 5D:
\beq
V^{(\infty)}=l\int\frac{d^4p}{(2\pi)^4}\sqrt{p^2+m^2(\ph)}
\eeq
The integral over $d^4p$ is obviously divergent, let us adopt regularization by a cut-off 
(as it was done in ref.~\cite{Delgado:1998qr})  and for comparison also the dimensional 
regularization:
\beq
V^{(\infty)}=\frac{R}{60\pi}\left\{
\baa{ll}
m^5(\ph)+
\frac12\sqrt{\lam^2+\ms}\left[3\lam^4+\lam^2\ms -2m^4(\ph)\right]
&{\rm cut-off}\non\\
m^5(\ph)&{\rm dim}\non
\eaa
\right.
\label{vinftyres}
\eeq
It is seen that $V^{(\infty)}$ is finite when the dimensional regularization is adopted.

As we have already mentioned there are two sources of divergences: the sum and the $d^4p$ integral.
In ref.~\cite{Delgado:1998qr} 
the sum was regularized-renormalized through the  differentiation and then integration with respect to
$E$, while for the divergent integral the result is shown in (\ref{vinftyres}) as the cut-off
option.
It turns out that the dimensional regularization of both the sum and the integral
proposed in ref.~\cite{GrootNibbelink:2001bx} leads to the same result as 
the one presented  above provided the integral is dimensionally regularized.
It will be instructive to compare both methods in order to understand the
puzzling agreement. In ref.~\cite{GrootNibbelink:2001bx} the sum is regularized by the 
following replacement (see eq.(11) of ref.~\cite{GrootNibbelink:2001bx}):
\beq
I=\int d^4p_4\sum_{n\geq 0}f(p_4,\frac{n}{R})\to\frac{1}{2\pi i}\int d^{D_4}p_4
\int_{\ominus}d^{D_5}p_5{\cal P}^+(p_5)f(p_4,p_5)\,,
\label{gn}
\eeq
where the notation of ref.~\cite{GrootNibbelink:2001bx} was adopted.
Then the author concludes that in fact it would be enough to regularize the integral since the divergent
part appears to be a function of $D_4+D_5$  only. 
For a first sight this statement looks confusing since we 
might have started with a divergent sum on the lhs of eq.(\ref{gn}). The sum is replaced 
by the integral over $d^{D_5}p_5$ and it looks that this regularization of the sum is needed.
The solution of this illusive puzzle seems to be the following. 
Note that for the effective potential calculation, the function $f(p_4,\frac{n}{R})$ depends on the
background field dependent mass $m(\ph)$ only through $p_4^2+m^2(\ph)$. Therefore a constant
that is $p_4$-independent on the lhs of (\ref{gn}) does not depend on $m(\ph)$ as well.
Since the divergence of the sum was dropped in the DPQ approach by the 
differentiation and then integration with respect to $E$ therefore we know that it 
was $p_4^2+m^2(\ph)$-independent constant. 
Let us now locate this divergence in the dimensional
approach. It turns out that it is hidden (and then erased)
in eq.(\ref{gn}), namely {\it the dimensional regularization 
of the integral over $d^4p_4$ at the same time regularize the integral and also
removes the constant ($p_4$-independent)
contribution to the sum}! This happens because of the following peculiar property of the 
dimensional regularization
\beq
\int d^{D_4}p_4 ({\rm constant}) =0\,.
\eeq
Therefore, no wonder that in fact it is not necessary
to regularize the sum if the dimensional regularization 
is adopted for the $d^4p_4$! The dimensional regularization takes care of both the divergent integral and
the divergent constant contribution to the sum.
So, it is clear now why both the method adopted by DPQ~\cite{Delgado:1998qr}  
and the one developed in 
ref.~\cite{GrootNibbelink:2001bx} lead to the same result\footnote{At most they may differ by
$m(\ph)$-independent constant. We have confirmed that by explicate calculation. The results are
identical separately for boson and fermion contributions to the effective potential.}.

In the remaining part of this paper we will apply methods developed in this section to 5D U(1) model
of universal extra dimensions. Then, expressions for the effective potential will either contain sums that
start at a zero mode ($n=0$) or at $n=1$\footnote{Note that in ref.~\cite{Delgado:1998qr} 
the summation is performed form $n=-\infty$ to $n=+\infty$, while here we have considered
separately the zero-mode contribution and the remaining KK modes from
$n=1$ to $n=+\infty$, that explains the factor $1/2$ in eq.(\ref{effpot}).}. Therefore the 
final result (for $\omega=0$) for both cases in 
dimensional regularization of the $d^4p$ integral
is the following :
\beq
V(m^2)=\frac12\left(V^{(\infty)}(m^2)+V^{(R)}(m^2) \pm V_0(m^2)\right)\,,
\label{effpot}
\eeq
where $+$ or $-$ corresponds to the zero mode included or excluded in the sum, respectively.
The contributions to the effective potential read:
\bea
V^{(\infty)}(m^2)&=&\frac{\pi R}{16\pi^2}\frac{4}{15}m^5(\ph)
\non\\
V_0(m^2)&=&\frac{1}{64\pi^2}\m4\left\{-C_{UV} +\ln\left(\frac{\ms}{\kappa^2}\right)
-\frac32\right\}\non\\
V^{(R)}(m^2)&=&-\frac{1}{64\pi^6}\frac{1}{R^4}
\left\{x^2{\rm Li}_3(e^{-x})+3x{\rm Li}_4(e^{-x})+3{\rm Li}_5(e^{-x})
\right\}\,,
\label{veffmsbar}
\eea
where $x\equiv2\pi R \sqrt{m^2(\ph)}$, $C_{UV}=\frac{2}{4-n}-\gamma_E +\ln(4\pi)$ 
($\gamma_E=0.5772\dots$ is the Euler-Mascheroni constant),
$\kappa$ is the regularization scale and
$V^{(\infty)}$ corresponds to the decompactification limit ($R \to \infty$), $V^{(R)}$ is the contribution 
from all the  KK modes (summed from $-\infty$ to $+\infty$)
and $V_0$ is the zero mode effective potential. 
The polylogarithm $Li_n(x)$ is defined by
\beq
Li_n(x)=\sum_{s=1}^\infty\frac{x^s}{s^n}\,.
\eeq
Note that $V^{(\infty)}$, $V_0$ and $V^{(R)}$ contributions
correspond exactly to the three terms separated
in ref.~\cite{GrootNibbelink:2001bx}  and denoted 
by $I_{5\,D}$, $I_{4\,D}$ and $I_{finite}$, respectively.
It is amazing that the divergence from the zero mode is still there, while in DPQ approach with
dimensional regularization it was gone (note that there the KK summation started at $n=-\infty$). 
This means that the singular contribution from the zero mode must be
canceled by the sum over $n\neq 0$ in the DPQ method. 
The explanation of this is the presence of the zero mode in the above consideration.

In order to get rid of the singularities present in $V_0$ we will adopt the $\overline{\rm MS}$
renormalization, then the 1-loop  contribution to the effective potential reads:
\beq
V_{eff\, ren}^{1-loop\; \overline{MS}}=\frac12\left(V^{(\infty)}(m^2)+V^{(R)}(m^2)
\pm V_{0\, finite}(m^2)\right)\,,
\label{veffrmsbaren}
\eeq
where $V_{0\, finite}$ is $V_{0}$ with the term $\propto C_{UV}$ subtracted.

\subsection{Decoupling of heavy KK modes.}
In sec.\ref{cut-off} we have discussed the one-loop effective potential
for 5D $\ph^4$ theory described by the tree-level potential specified in eq.(\ref{vphi4}).
Using the cut-off regularization of the 4D integral and 
adopting the on-shell renormalization conditions (\ref{rencon})
we have found in eq.(\ref{effpotren1})
the renormalized effective potential that originates from the first $n_{max}$ KK modes 
that can be written as:
\beq
V_{eff\, ren\;I}^{1-loop\; \lam}= \sum_{n=0}^{n_{\max}}V_n(\ph)
\eeq
for
\beq
V_n(\ph)=
\frac{1}{32\pi^2}\left\{\frac12\left(\ms+\mn2\right)^2\ln\left(\frac{\ms+\mn2}{m^2(0)+\mn2}\right)
-\frac34 m^4(\ph) + m^2(\ph)\left[m^2(0)-\frac12\mn2\right]\right\}\,,
\label{vn}
\eeq
where $m^2(\ph)$ was defined in eq.(\ref{m2}).
In order to investigate the decoupling of heavy KK modes (corresponding to large $n$)
in the model 
it is useful to expand $V_n(\ph)$ in the limit of $n\to\infty$ and then sum over $n$:
\beq
V_{eff\, ren\;I}^{1-loop\; \lam}= \frac{1}{32\pi^2}\sum_{n=0}^{n_{\max}}\left[
-\frac12 m^2(0)\mn2 + \frac14 m^4(0) + {\cal O}\left(\frac{1}{n^2}\right)\right]
\label{vnlim}
\eeq
As it is seen, only leading ($\sim \mn2$) and sub-leading ($m_n$-independent) terms are
divergent when the summation over $n$ is performed in the limit $\nm\to\infty$. 
The key observations is that
those terms are $\phi$ independent! 
Even though the above sum is divergent, the divergence 
is a constant, $\ph$-independent contribution to the effective potential
and therefore will be irrelevant.
That happens because there is no couplings that could grow with 
$n$\footnote{In the next section we will discuss in details the 5D model based on U(1)
gauge symmetry. We will observe there that mass matrix for the $(A_{5\,n},\chi_n)$ system is 
non-diagonal and in fact the off-diagonal entries are of the form $n\ph/R$, so that
suggest that there exist coupling constants growing with $n$. However
as it will be seen, the determinant and the trace of the mass matrix grows as
$n^4/R^4$ and $n^2/R^2$, therefore even in that case in the limit of large $n$
we shall anticipate decoupling of heavy modes. The explicit calculations confirm
this expectation.}. Of course, the remaining, finite part of the effective potential (denoted in 
eq.(\ref{vnlim}) by ${\cal O}(1/n^2)$) depends on $\ph$ and leads to the 
genuine effective potential\footnote{The corresponding analogous phenomena could be 
also found in the method 
of DPQ~\cite{Delgado:1998qr}; as it was discussed earlier,  an infinite $\ph$-independent
term was dropped there through differentiation and subsequent integration over $E$.}.
In other words, the decoupling of heavy KK modes takes place as a consequence of
renormalization of the cosmological constant.

\begin{figure}[htb]
\psfull
\begin{center}
  \leavevmode
\epsfig{file=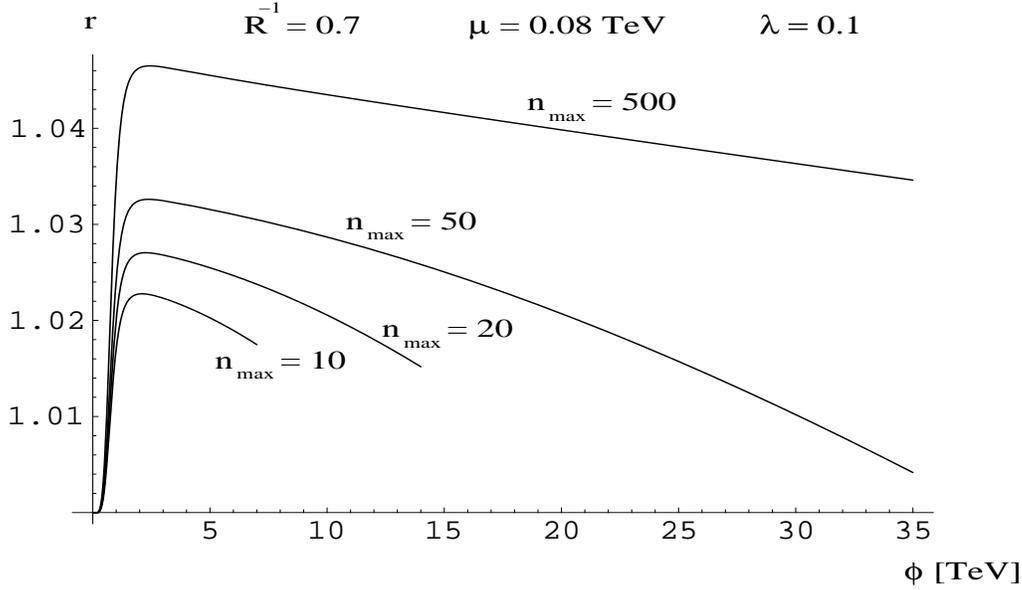,height=8.cm,width=14cm
}
\end{center}
\caption{The ratio defined by eq.(\ref{xratio}) for the $\ph^4$ theory
for $\kappa=n_{max}/(2R)$, $R=1/0.7$, $\mu=0.08\tev$, $\lambda=0.1$, and the shift parameter 
$V_0=0.01 \tev^4$. The curves
from the left to the right correspond to increasing cut-offs: $n_{max}=10, 20,50$ and $500$.}
\label{ratio}
\end{figure}

In order to discuss the decoupling more quantitatively, it is worth to 
compare the effective potential obtained within the cut-off regularization 
(\ref{effpotren1}) with the one for the minimal subtraction (\ref{veffrmsbaren}).
One could wish to plot the simple ratio: 
$V_{eff\, ren\;I}^{1-loop\; \lam}(\ph)/V_{eff\, ren}^{1-loop\; \overline{MS}}(\ph)$.
However, it turns out that in the vicinity of $\ph\simeq 0$ the $\overline{\rm MS}$-renormalized
1-loop contribution to the effective 
potential has a zero and the plot of the ratio is very unstable. Fortunately, the value of 1-loop
contribution both to $V_{eff\, ren\;I}^{1-loop\; \lam}(\ph)$ and 
$V_{eff\, ren}^{1-loop\; \overline{MS}}(\ph)$ is in this region
by far negligible comparing to the tree
level contribution. Therefore we will modify the naive ratio as follows:
\bit 
\item Since the tree-level potential 
is the reference point for 1-loop corrections therefore we will add $V_{tree}(\ph)$ 
both in the denominator and the numerator. 
\item To eliminate the unwanted irrelevant constant contributions\footnote{It is especially 
important in light of proceeding discussion of the decoupling in the case of 
the cut-off regularization.} to the effective potentials we will subtract $V_{eff}(0)$ contributions
both in the denominator and the numerator. 
\item The effective potentials obtained according to the above prescription have
zeros in the vicinity $\ph\simeq 0$ that are slightly misplaced in the denominator 
and the numerator, therefore we introduce a constant shift $V_0$ in order 
to screen the instability caused by the zero of the denominator.
\eit
So, we will adopt the following ratio to compare the cut-off and 
dimensional regularization:

\beq
r(\kappa,n_{max};\ph)\equiv 
\frac{V_{tree}(\ph)+\left[V_{eff\, ren\;I}^{1-loop\; \lam}(\ph)-V_{eff\, ren\;I}^{1-loop\; \lam}(0)\right]+V_0}
{V_{tree}(\ph)+\left[V_{eff\, ren}^{1-loop\; \overline{MS}}(\ph)-V_{eff\, ren}^{1-loop\; \overline{MS}}(0)\right]
+V_0}
\label{xratio}
\eeq
The ratio $r=r(\kappa,n_{max};\ph)$ is, of course, a function of the cut-off ($\Lambda_5=n_{max}/R$) and
the regularization scale ($\kappa$). In fig.\ref{ratio} we plot $r=r(n_{max}/(2R), n_{max};\ph)$
for $n_{max}=10, 20, 50$ and $500$, what corresponds to the choice\footnote{Other possible
choices of $\kappa$, e.g. $\kappa=\ph$, do not change results for $r$ substantially.
Note, that here we have decided to adopt the same cut-off for 4D and 5D: $\Lambda_5=\Lambda$.} of the 
regularization scale $\kappa=\Lambda/2$.
For a given $\nm$, the ratio  $r=r(n_{max}/(2R), n_{max};\ph)$ is plotted against $\ph$ varying from 0 
up to the appropriate cut-off $\Lambda=n_{max}/R$. Note, however, that the cut-off 
corresponding to $\nm=500$, $\Lambda=350\tev$, is not shown for the sake of clarity of the figure. However,
it has been checked that even in this case $r$ remains within the 5\% distance from 1.
For the purpose of fig.\ref{ratio}, we have used the mass parameter $\mu=0.08 \tev$, 
the quartic coupling constant $\lambda=0.1$, and the shift $V_0=0.01 \tev^4$~\footnote{If we plotted 
$r$ for $\ph\gsim1\tev$ (that is large enough to pass the zero of the denominator) 
we would not need to introduce $V_0$.}. It has been checked that for $0\leq V_0\leq 1\tev^4$ the ratio
$r$ remains below $1.05$ for $\ph\gsim1\tev$ even though the shape in the region $1\lsim\ph\lsim5\tev$
is influenced by the choice of $V_0$. However, it should be emphasized that for $\ph \gsim 5 \tev$ 
(for the stability we will discuss the effective potential for field strength $\ph \gg 1\tev$)
the curves are almost insensitive to $V_0$. 

As it is seen from the plot, even though for small $\ph$, $r(\kappa,\nm;\ph)$ is a monotonically rising
function of $\nm$ (the curves corresponding to growing $\nm$ are being shifted up), nevertheless, 
eventually for larger $\ph$, $r$ approaches 1 closer for curves corresponding to larger $\nm$. 
In fact, this is what we should expect if the effective potential calculated in the cut-off and 
$\overline{MS}$ schemes were close.

Conclusion that can be drawn from this picture is that 
the cut-off and the minimal subtraction schemes are consistent
and the dependence on the cut-off is very weak. One should however remember 
that we have adopted two different 
renormalization schemes and therefore the agreement is never expected to be perfect.
%

\section{U(1) Model}
\label{u1model}

In this section we will construct a simple 5D model that could successfully mimic the SM
as far as the shape of the effective potential is concerned. For a gauge group we  choose 
U(1). In order to break spontaneously the symmetry, we will introduce a complex scalar $\phi$.
To  have a zero-mode massive fermion (the analog of the top quark) we will have to introduce
two 5D fermions: $\psi$ and $\lambda$.
The model is defined by the Lagrangian density:

\beq
\lcal(x,y)=-\frac14 F^{MN}F_{MN}+(D_M\phi )^\star (D^M\phi)-V^{(5)}(\phi)+
\lcal_{GF}^{(5)}+\lcal_f^{(5)}\,,
\label{lag}
\eeq
where
\bea
F_{MN}(x,y)&\equiv&\partial_MA_N(x,y)-\partial_NA_M(x,y)\non\\
D_M&\equiv& \partial_M +ie_5 A_M(x,y)\non\\
V^{(5)}(\phi)&\equiv&\mu^2|\phi|^2+\lambda_5|\phi|^4\non\\
\phi(x,y)&=&\frac{1}{\sqrt{2}}\left[h(x,y)+i\chi (x,y)\right]\non\\
y\equiv x^4\,,\non
\eea
where $A_M$ is a gauge field, $D_M$ is a covariant derivative. We will assume that the 
tree-level potential is stable, so $\lambda_5>0$.

Hereafter  we will adopt the following form of the gauge 
fixing Lagrangian\footnote{For discussion of the Lorentz non-covariant $R_\xi$ gauges,  
see refs.~\cite{Muck:2002af},\cite{Muck:2001yv}.}:
\beq
\lcal_{GF}^{(5)}=-\frac{1}{2\xi}\left[\partial_\mu A^\mu-\xi\left(\partial_5A_5+
e_5\frac{v\chi}{\sqrt{2\pi R}}\right)\right]^2\, ,
\eeq
where $v=\langle h_0\rangle$ is the vacuum expectation value of the zero mode 
of the scalar $h(x,y)$.

In order to generate massive zero-modes for fermions we will introduce here two fermion 
fields, one 
charged ($\psi(x,y)$) and one neutral ($\lambda(x,y)$) 
under U(1):

\bea
\lcal^{(5)}_f&=&\bar{\psi}(x,y)\gamma^M\left[i\partial_M+e_5A_M\right]\psi(x,y)+
\bar{\lambda}(x,y)\gamma^Mi\partial_M\lambda(x,y)+\non\\
&&-\left[g_5\bar{\psi}(x,y)\phi(x,y)\lambda(x,y) + \hc\right]\,.
\label{lagfer}
\eea

The action of the U(1) local symmetry  is defined by:
\bea
\phi(x,y) &\to& e^{-ie_5\theta(x,y)}\phi(x,y)\non\\
\psi(x,y) &\to& e^{-ie_5\theta(x,y)}\psi(x,y)\non\\
\lambda(x,y) &\to& \lambda(x,y)\non\\
A_M(x,y) &\to& A_M(x,y)+\partial_M\theta(x,y)\,.
\label{symm}
\eea

The compactification of the extra dimension is specified by the following
$S^1/Z_2$ orbifold conditions:\\
$\bullet$  all the fields and the gauge function $\theta(x,y)$ remain unchanged under 
a shift $y\to y+2\pi R$,\\
$\bullet$ 
\vspace{-1cm}
\bea
&A_\mu(x,y)=A_\mu(x,-y) \lsp &A_5(x,y)=-A_5(x,-y)\non\\
&\phi(x,y)=\phi(x,-y)&\non\\
&\psi_R(x,y)=\psi_R(x,-y) \lsp & \psi_L(x,y)=-\psi_L(x,-y)\non\\
&\lambda_L(x,y)=\lambda_L(x,-y) \lsp &\lambda_R(x,y)=-\lambda_R(x,-y)\non\\
&\theta(x,y)=\theta(x,-y)\,.&
\label{orbi}
\eea

KK expansions read:
\bea
A^\mu(x,y)&=&\frac{1}{\sqrt{2\pi R}}\left[A^\mu_0(x)+
\sq2\sum_{n=1}^{\infty}A^\mu_n(x)\cos\arg\right]\non\\
A^5(x,y)&=&\frac{1}{\sqrt{\pi R}}\sum_{n=0}^{\infty}A^5_n(x)\sin\arg \non\\
\phi(x,y)&=&\frac{1}{\sqrt{\pi R}}\sum_{n=0}^{\infty}\phi_n(x)\cos\arg\\
\psi(x,y)&=&\frac{1}{\sqrt{2\pi R}}\left\{\psi_{R\,0}(x)+
\sq2\sum_{n=1}^{\infty}\left[\psi_{R\,n}(x)\cos\arg + \psi_{L\,n}(x)\sin\arg\right]\right\}\non\\
\lambda(x,y)&=&\frac{1}{\sqrt{2\pi R}}\left\{\lambda_{L\,0}(x)+
\sq2\sum_{n=1}^{\infty}\left[\lambda_{L\,n}(x)\cos\arg + \lambda_{R\,n}(x)\sin\arg\right]\right\}\non\\
\theta(x,y)&=&\frac{1}{\sqrt{\pi R}}\sum_{n=0}^{\infty}\theta_n(x)\cos\arg
\non\,,
\label{kk}
\eea
where $m_n\equiv n/R$, subscripts R and L are referring to 4D chiral fields and it is assumed that
$A_{5\, 0}=0$. In the following we will adopt
the following notation for the real and imaginary parts of $\phi_n(x)$:
\beq
\phi_0=\frac12\left(h_0+i\chi_0\right), \lsp \phi_{n\neq 0}=\frac{1}{\sqrt{2}}(h_n+i\chi_n)\, .
\eeq

It is worth noticing that after compactification the 4D Lagrangian
expressed in terms of KK modes is still gauge invariant and
the $U(1)$ transformations of the gauge fields  read:
\bea
A_{n\,\mu}(x) & \to & 
\left\{
\baa{lll}
A_{0\,\mu}(x) + \sqrt{2}\partial_\mu\theta_0(x) & {\rm for} &n=0 \\
A_{n\,\mu}(x) + \partial_\mu\theta_n(x) & {\rm for} &n\neq 0
\eaa
\right.\label{gaugeAmu}\\
A_{n\,5}(x) & \to & A_{n\,5}(x) -\frac{n}{R}\theta_n(x)\,.\label{gaugeA5}
\eea
The corresponding infinitesimal transformation for $\phi_n(x)$ is the following:
\beq
\baa{l}
\phi_{0}(x) \to \phi_{0}(x)-\frac{ie}{\sqrt{2}} ( 2 \theta_{0}(x)\phi_{0}(x)+\sum_{m=1}^{\infty}\theta_{m}(x)\phi_{m}(x) ), \\
\phi_n(x) \to \phi_n(x)-\frac{ie}{\sqrt{2}}\sum_{m,l=0}^{\infty}A_{nml}\theta_m(x)\phi_l(x)\,,  
\eaa
\label{gaugephi}
\eeq
where $A_{nml}$ is defined in the Appendix A and $e \equiv e_5/\sqrt{2\pi R}$.

The goal of this paper is to investigate stability of the ground state of the model.
Therefore first we have to determine the tree level potential, the next step will be to
calculate the effective potential at the 1-loop level.
Expanding in KK modes and integrating over $y$ yields the following 4D potential:
\bea
V^{(4)}&=&\sum_{n=0}^{\infty}\left[m_n^2+\mu^2\right]\phi_n^\star \phi_n+
\mu^2 \phi_0^\star\phi_0 +
\frac{\lambda}{2} \sum_{n,m,k,l=0}^{\infty}B_{nmkl}\phi_n^\star \phi_m \phi_k^\star \phi_l + \non\\
&&\frac{e^2}{2}\sum_{n,m,k,l=0}^{\infty}D_{nmkl}A_{5\,n}A_{5\,m}\phi_k^\star \phi_l - 
\frac{ie}{\sqrt{2}}\sum_{n,m,k=0}^{\infty}C_{nmk}m_nA_{5\,m}(\phi_n^\star\phi_k-\phi_k^\star\phi_n) +\non\\
&&\frac{\xi}{2}\sum_{n=0}^{\infty}\left(m_nA_{5\,n}+
ve\chi_n\right)^2 \,,
\label{4dpot}
\eea
where $\lambda\equiv \lambda_5/(2\pi R)$ and the coefficients
$B_{nmkl},D_{nmkl}$ and $C_{nmk}$ are defined in the Appendix A.

In spite of the fact that the potential looks complicated, it is easy to see that for 
$\lambda_5>0$ the potential is
positive definite in the limit of $|\phi_n|^2\to \infty$ and therefore the ground state is stable.
The 4D potential emerges from the 5D potential, the Higgs-boson kinetic term and the
gauge fixing term:
\beq
V^{(4)}=\int_o^{2\pi R}dy\;\left[V_5(x,y)+(D_5\phi)^\star(D_5\phi)+
\frac{\xi}{2}\left(\partial_5A_5+
e_5\frac{v\chi}{\sqrt{2\pi R}}\right)^2\right]\,,
\label{4dpotder}
\eeq
where $D_5\phi$ is the fifth component of the covariant derivative of the Higgs field and
the last term emerges from the gauge fixing term.
So, it is clear that the 4D potential must be positive definite as it is an integral 
over a positive function. In the following we will investigate 1-loop corrections to the
effective potential.

We will consider the  case $\mu^2<0$, then it is easy to see that if $-\mu^2\leq 1/R^2$ then only
the zero mode $h_0(x)$ can develop a non-zero vacuum expectation value, at the tree level 
we get:
\beq
\langle h_0(x) \rangle \equiv v=\sqrt{\frac{-\mu^2}{\lambda}}\,.
\label{vev}
\eeq

We will calculate  the effective potential in the direction of the tree level vacuum:
$\chi_0=h_n=\chi_n=A_{5\, n}=0$ and $h_0\neq 0$.
The Landau gauge defined here by $\xi=0$ will be adopted hereafter.

We will expand the 4D Lagrangian around $\chi_0=h_n=\chi_n=A_{5\, n}=0$ and $h_0\to h_0+\ph$,
where $\ph$ is the classical constant (in 4D) external background field for the calculation of one-loop
Green's functions that are necessary for the effective potential. Then in the Landau gauge the
following mass terms are obtained:
\bea
m_{h_0}^2\equiv\frac{\partial^2V}{\partial h_0^2}&=&\mu^2+3\lambda \ph^2\non\\
m_{\chi_0}^2\equiv\frac{\partial^2V}{\partial \chi_0^2}&=&\mu^2+\lambda \ph^2\non\\
m_{h_n\,h_m}^2\equiv\frac{\partial^2V}{\partial h_n \partial h_m }&=&
(m_n^2+ 
\mu^2 + 3\lambda \ph^2)\delta_{nm}\equiv m_{h_n}^2\delta_{nm}\non\\
\left[
\baa{cc}
\frac{\partial^2V}{\partial A_{5\,n} \partial A_{5\, m}}&
\frac{\partial^2V}{\partial A_{5\,n} \partial \chi_m}\\
\frac{\partial^2V}{\partial A_{5\,m} \partial \chi_n}&
\frac{\partial^2V}{\partial  \chi_n \partial \chi_m}\\
\eaa
\right]
&=&
\left[
\baa{cc}
e^2\ph^2&-em_n\ph\\
-em_n\ph& (m_n^2 +\mu^2+\lambda\ph^2)\\
\eaa
\right]\delta_{nm}\non
\label{masses}
\eea

In the following part of this section we will show separate contributions to the effective 
potential calculated in the  $\overline{\rm MS}$ scheme in dimensional regularization.

Let us start with the $(A_{5\,n},\chi_m)$ system.
The mixing in the mass matrix for $A_{5\,n}$ and $\chi_m$ causes some technical difficulties
that are described in Appendix B. The final result for the $(A_{5\,n},\chi_m)$ system is the
following:
\beq
V_{eff}^{(A_5,\chi)}=\frac12\left(V^{(\infty)}_{mix}+V^{(R)}_{mix}-V_{0\, finite}^{(A_0)}
-V_{0\, finite}^{(\chi_0)}\right)\,,
\label{eff_mix}
\eeq
where $V^{(\infty)}_{mix}$ and $V^{(R)}_{mix}$ are the analogs of the ``divergent'' 
and finite contributions  to the effective potential (\ref{veffmsbar}) in the case of mixing:
\bea
V^{(\infty)}_{mix}&=&-\frac{y^{1/2}(y^2-1)x^5}{2^{12}\sqrt{2}\pi^5R^4}
{\rm F}\left(-\frac14,\frac74;2;1-\frac{1}{y^2}\right) \\
V^{(R)}_{mix}&=& -\frac{y^{3/2}(1+y)^{1/4}x^{7/2}}{2^9\pi^5\sqrt{\pi}R^4}
{\rm Li}_{\frac32}\left(e^{-x\sqrt{1+y}}\right)\,,
\eea
where ${\rm F}(a,b;c;z)$ is the  hypergeometric function,
\beq
x\equiv 2\pi R \sqrt{a} \lsp {\rm and} \lsp y\equiv \frac{2\sqrt{b}}{a}
\label{xy}
\eeq
for $a$ and $b$  defined in eq.(\ref{ab}).
$V_{0\, finite}^{(A_0)}$ and $V_{0\, finite}^{(\chi_0)}$
are the finite parts of scalar contributions (see eq.(\ref{veffmsbar})) to the 
effective potential calculated for 
the zero mode vector boson mass ($m_{A_0}^2=e^2\ph^2$)
and Goldstone boson ($m_{\chi_0}^2=\mu^2+\lambda \ph^2$), respectively.

All neutral scalar modes contribute to the effective potential as follows:
\beq
V_{eff}^{(s)}(\ph)=
\frac12\left[V^{(\infty)}(m_{h_0}^2)+V^{(R)}(m_{h_0}^2)+V_{0\, finite}(m_{h_0}^2)\right]+
V_{0\, finite}(m_{\chi_0}^2)+V_{eff}^{(A_5,\chi)}(\ph)\,.
\label{scalpot}
\eeq
For the vector boson contribution we get
\beq
V_{eff}^{(v)}(\ph)=
\frac32\left[V^{(\infty)}(m_{A_0}^2)+V^{(R)}(m_{A_0}^2)+V_{0\, finite}^v(m_{A_0}^2)\right]\,,
\label{vecpot}
\eeq
where the zero-mode vector contribution reads 
\beq
V_0^v(m^2)=\frac{1}{64\pi^2}\m4\left\{-C_{UV} +\ln\left(\frac{\ms}{\kappa^2}\right)
-\frac56\right\}\,,
\eeq
and  $m_{A_n}^2=m_n^2 + e^2\ph^2$.

After KK expansion and integration over $y$ the 4D fermionic Lagrangian reads 
(see ref.~\cite{Papavassiliou:2001be} 
for a similar construction):
\beq
\lcal_f^{(4)}=\bar{f_0}(i\gamma^\mu \partial_\mu - m_{f\,0}) f_0+
\sum_{n=1}^{\infty}\left[\bar{\xi_n}i\gamma^\mu \partial_\mu \xi_n- \bar{\xi_n}M_n\xi_n\right]\,,
\label{lfermi}
\eeq
where $m_{f_0}=g\ph/\sq2$, $g\equiv g_5/\sqrt{2\pi R}$, $m_n=n/R$ and
\beq
f_0=\psi_{R\,0}+\lambda_{L\,0} \lsp 
\xi_n=
\left(
\baa{c}
\psi_{R\,n}+\psi_{L\,n}\\
\lambda_{R\,n}+\lambda_{L\,n}
\eaa
\right)
\lsp
M_n=
\left(
\baa{cc}
-m_n&m_{f_0}\\
m_{f_0}&+m_n
\eaa
\right)
\eeq
After diagonalization the fermionic mass matrix reads:
\beq
{\cal M}=\pm
\left(
\baa{cc}
-(m_n^2+m_{f_0}^2)^{1/2}&0\\
0&(m_n^2+m_{f_0}^2)^{1/2}
\eaa
\right)
\eeq
So, we have two fermions degenerate in masses (the minus in front of the upper component mass 
can be removed through a chiral rotation).

Fermions (no color degrees of freedom included) contribute to the effective potential as follows:
\beq
V_{eff}^{(f)}(\ph)=
-4V_{0\, finite}(m_{f_0}^2)
-\frac82\left[V^{(\infty)}(m_{f_0}^2)+V^{(R)}(m_{f_0}^2)- V_{0\, finite}(m_{f_0}^2)\right]\,,
\label{fermpot}
\eeq

\begin{figure}[htb]
\psfull
\begin{center}
  \leavevmode
\epsfig{file=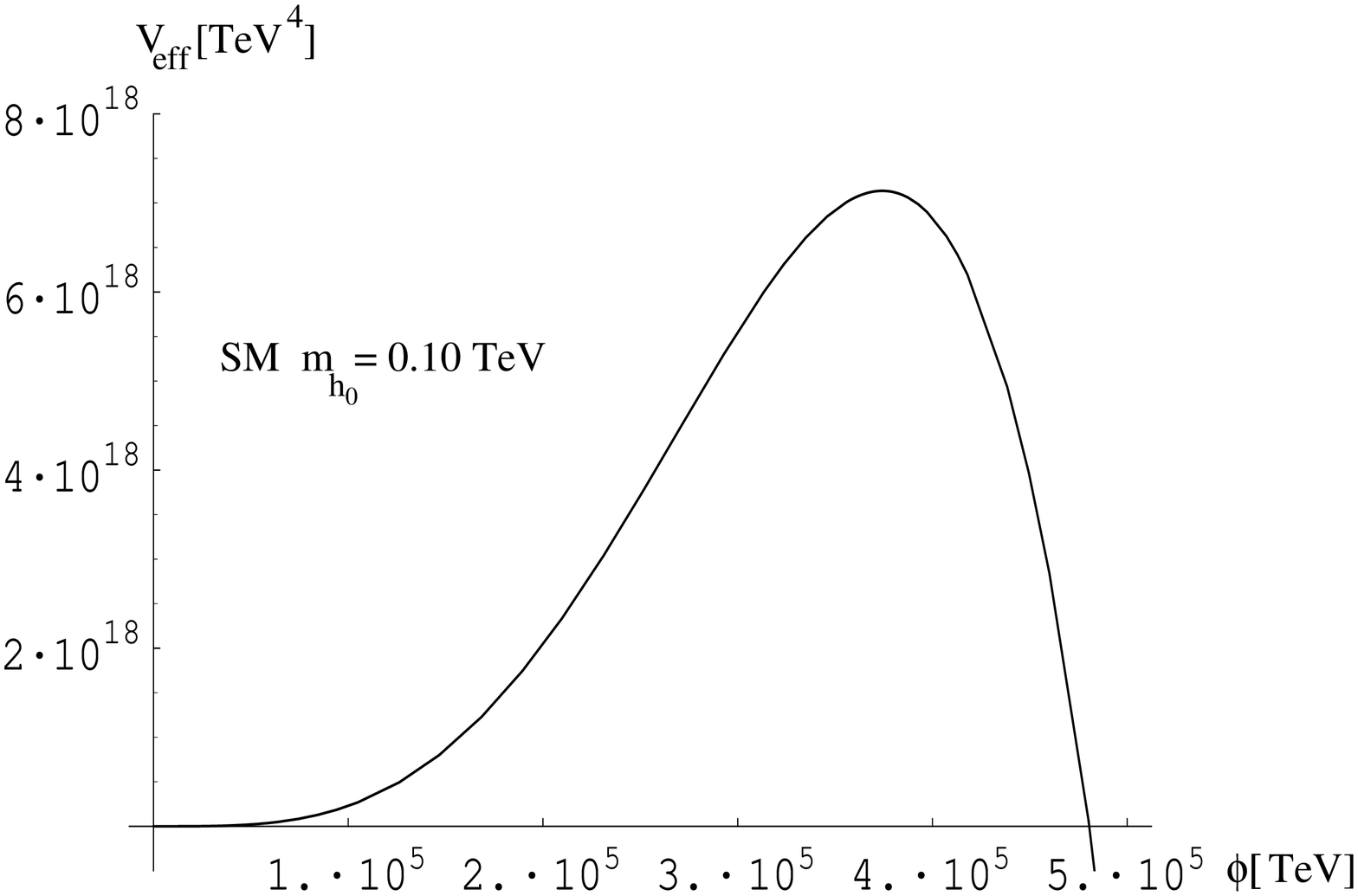,height=6.cm,width=14cm
}
 \end{center}
\caption{The zero-mode (SM-like) 1-loop effective potential for $m_{h_0}=0.10\tev$ in the dimensional 
regularization.}
\label{smeffpot}
\end{figure}
\begin{figure}[ht]
\psfull
\begin{center}
  \leavevmode
\epsfig{file=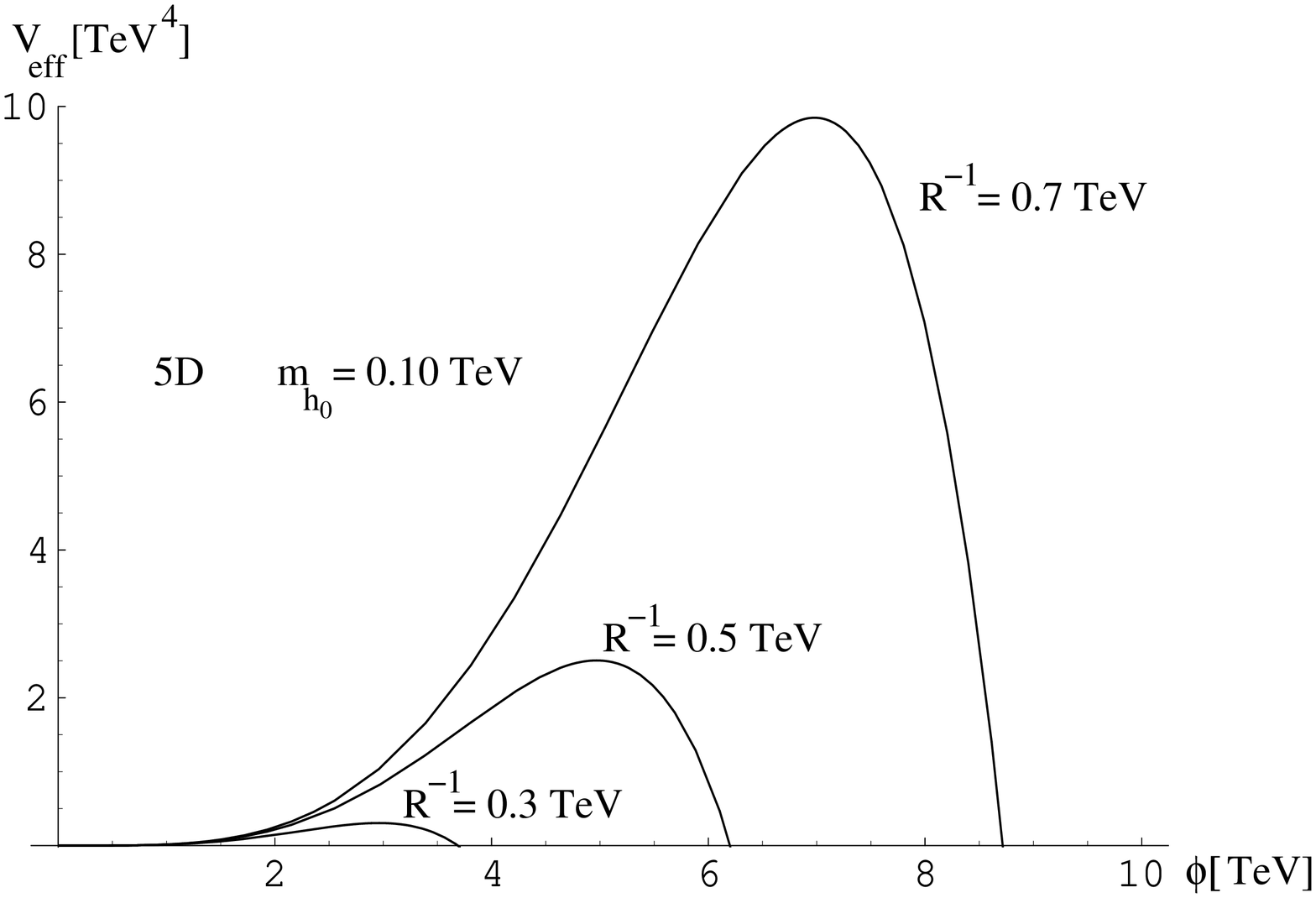,height=6.cm,width=14cm
}
 \end{center}
\caption{The full 1-loop effective potential in the dimensional regularization for $m_{h_0}=0.10\tev$.
The compactification radius $R^{-1}=0.3,0.5,0.7 \tev$ 
was adopted (higher curves correspond to smaller $R$).
All other parameters are specified in the text.}
\label{smeffpot}
\end{figure}

\noindent for $m_{f_n}^2=m_n^2+m_{f_0}^2$.

Eventually, the total 1-loop effective potential is given by the following formula:
\beq
V_{eff}^{1-loop}=V_{tree}+V_{eff}^{(s)}+V_{eff}^{(v)}+V_{eff}^{(f)}\,,
\label{vtot}
\eeq
where 
\beq
V_{tree}(\ph)=\frac{\mu^2}{2}\ph^2+\frac{\lambda}{4}\ph^4\,.
\label{vtree}
\eeq

\section{Results}
\label{results}
In order to mimic the SM we have adopted the following parameters for the plots:
$e=\sqrt{4\pi/137}$, $v=0.246\tev$,
the fermion zero-mode mass $m_{f\,0}=0.150\tev$ and  the renormalization scale $\kappa= 0.1\tev$.
We will adopt the asymptotic formula for $V^{(R)}_{mix}$ given in eq.(\ref{mix_r_veff}),
however it should be emphasized that it
provides an excellent approximation in the whole parameter range that is of interest here.

It is seen from the plots that effects of non-zero KK modes are very dramatic. For instance,
for $m_{h_0}=0.10\tev$ and $R^{-1}=0.3\tev$ the instability scale is shifted down from
$4.8\times 10^5\tev$ to $3.6 \tev$ ! The model is much less stable as a consequence of
the presence of the KK modes. Closer inspection shows that the result
is triggered by the fermionic contribution to the 4D effective potential 
and the leading contribution emerges from $V^{(\infty)}$. Note that since 
we wished to construct a model that would posses a zero-mode massive fermion
therefore it was necessary to introduce the extra 5D fermion. As a consequence
the model contains after reduction to 4D doubly degenerated Dirac fermions 
for each KK mode what enhances the fermionic contributions and is the source
of the extra factor of 2 in front of the second term in eq.(\ref{fermpot}).
If the factor 2 is removed (just to test the effect of fermion doubling)
the result changes and for instance for $m_{h_0}=0.10\tev$ and $R^{-1}=0.3\tev$ 
the instability appears at $6.5\tev$ instead of $3.6 \tev$, obviously the model would 
be more stable. 
It turns out that for our model (with full spectrum of fermions)
the fermionic KK contribution is by factor of $2.5\div 5$ larger (for 
$\ph \simeq 0.5 \div 3.5 \tev$ at $m_{h_0}=0.10\tev$ and $R^{-1}=0.3\tev$) 
than the zero-mode contribution. As a consequence
the tree level potential bends down more rapidly for much lower field strengths than
for the zero mode only.

\section{Conclusions}
\label{summary}

We have discussed the effective potential in 4-dimensional  models 
that originate from 5-dimensional ones
reduced down to 4 dimensions.  The cut-off and the dimensional regularization
schemes were discussed and compared. It  was shown that the prescriptions are
consistent with each other and lead to the same physical consequences.
\begin{figure}[htb]
\psfull
\begin{center}
  \leavevmode
\epsfig{file=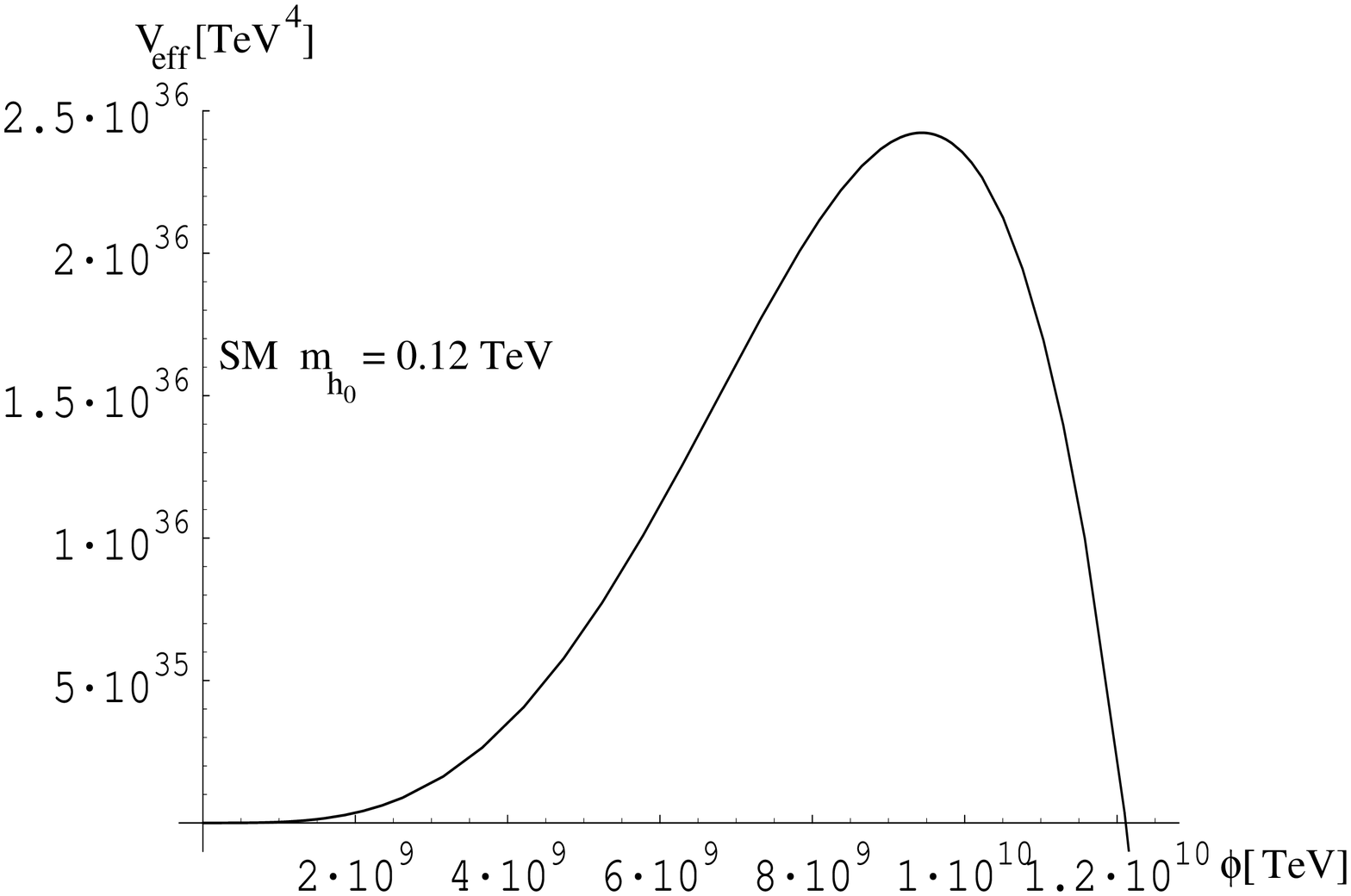,height=6.cm,width=14cm
}
 \end{center}
\caption{The zero-mode (SM-like) 1-loop effective potential for $m_{h_0}=0.12\tev$ in the dimensional 
regularization.}
\label{smeffpot}
\end{figure}
\begin{figure}[htb]
\psfull
\begin{center}
  \leavevmode
\epsfig{file=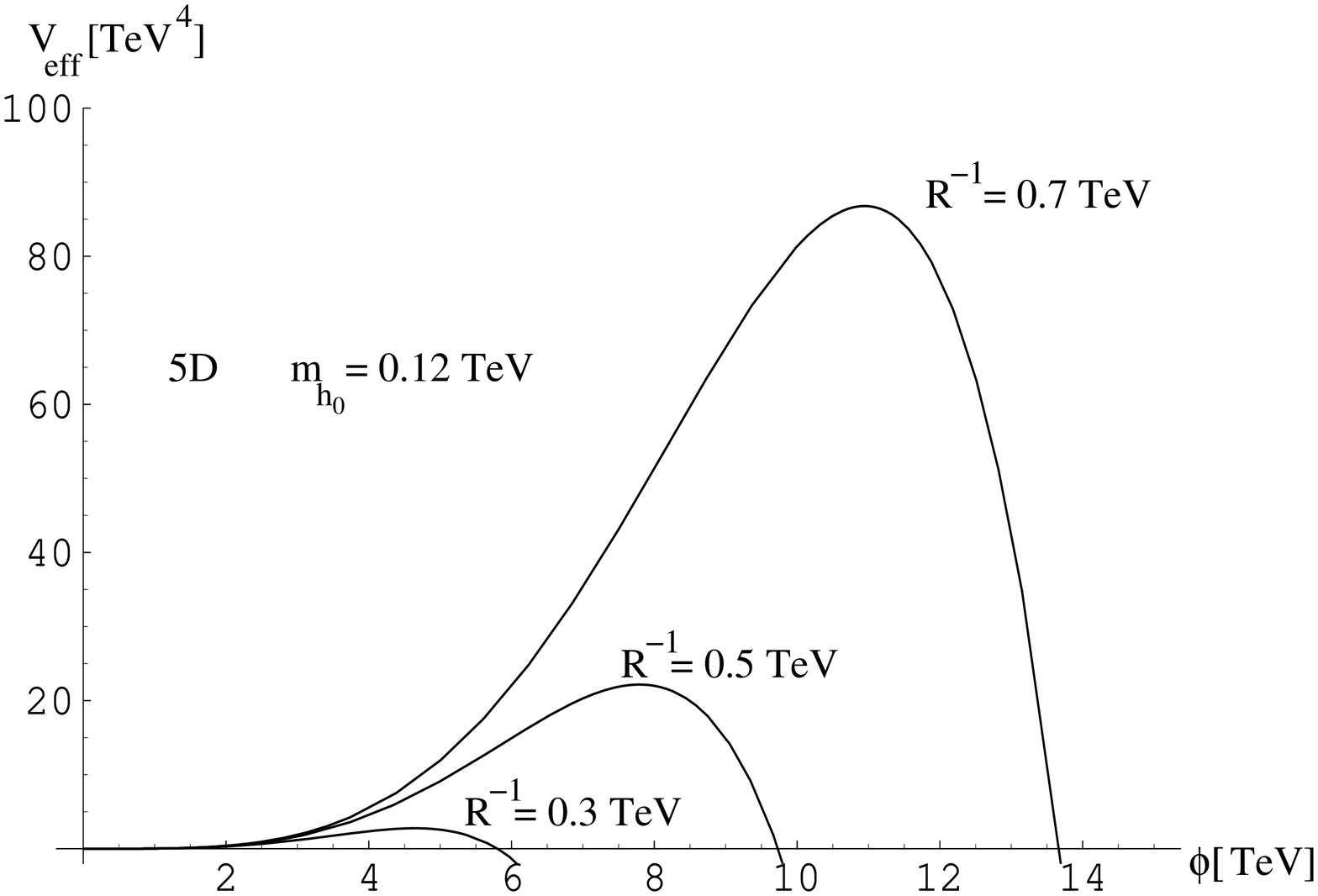,height=6.cm,width=14cm
}
 \end{center}
\caption{The full 1-loop effective potential in the dimensional regularization for $m_{h_0}=0.12\tev$.
The compactification radius $R^{-1}=0.3,0.5,0.7 \tev$ 
was adopted (higher curves correspond to smaller $R$).
All other parameters are specified in the text.}
\label{smeffpot}
\end{figure}
It turned out that when the number of KK modes included ($\nm$) varies between 10 and 500,
the effective potential calculated within the cut-off regularization accompanied by the on-shell renormalization
is never farther than $5\%$ away from the potential found in the  dimensional regularization
with $\overline{\rm MS}$. 

In order to take into account non-diagonal mass matrices
we have generalized the standard technique for the calculation of KK contributions
to the effective potential developed by Delgado, Pomarol and Quir\'os
in ref.~\cite{Delgado:1998qr}.
We have constructed a simple U(1) 5-dimensional model containing gauge boson,
a complex scalar and two fermions. The model parameters were adjusted, so that the model
should mimic 5-dimensional extension of the Standard Model. The one-loop
effective potential for the model was calculated adopting the dimensional regularization with the 
$\overline{\rm MS}$ renormalization. Like in the Standard Model the effective potential
turned out to be unbounded from below as a consequence of fermionic contributions.
It has been found that the presence of the tower of fermionic KK modes leads to a  major
modification of the effective potential and in particular could substantially lower the scale 
of instability. For instance,
for $m_{h_0}=0.10\tev$ and $R^{-1}=0.3\tev$ the instability scale is shifted down from
the Standard Model value 
$4.8\times 10^5\tev$ to $3.6 \tev$ ! The model is much less stable as a consequence of
the presence of the KK modes.
The same qualitative behavior of the effective potential is expected for the true 5-dimensional
extension of the Standard Model. The order of magnitude for the instability scale should not
differ very much from the results presented here,
however for a definite prediction for the instability scale as a function of 
the Higgs-boson mass a dedicated
study is necessary\cite{inprogress}.

\vspace*{2cm}

\centerline{ACKNOWLEDGMENTS}

The authors are very grateful to Jos\'e Wudka for his collaboration in early stages of this work.
They also thank  Adam Falkowski, Zygmunt Lalak,  Krzysztof Meissner,
Marek Olechowski and Jacek Pawelczyk for useful discussions. 
P.B. was supported by the EU fifth Framework Network ``Supersymmetry and the Early Universe''
(HPRN-CT-2000-00152).
B.G. was supported in part by the State Committee for Scientific
Research under grant 5~P03B~121~20 (Poland). 

\vspace*{2cm}

\renewcommand{\theequation}{A.\arabic{equation}}
\setcounter{equation}{0}
\noindent \hspace*{-0.72cm}
%
\centerline{\bf APPENDIX A}

The integrals used in the text:
\bea
&&\int_0^{2\pi R}\cos(m_ny)dy=(2 \pi R) \delta_{n,0}\non\\
&&\int_0^{2\pi R}\sin(m_ny)dy=0\non\\
&&\int_0^{2\pi R}\cos(m_ny)\cos(m_my)dy=\left\{ 
\baa{lll} 
(\pi R)\, \delta_{n,m} & {\rm for} & n,m\neq 0 \\
 2 \pi R & {\rm for} & n,m =0 
\eaa 
\right. \non\\
&&\int_0^{2\pi R}\sin(m_ny)\sin(m_my)dy=\left\{ 
\baa{lll} (\pi R)\, \delta_{n,m} & {\rm for} & n,m \neq 0\\ 
0 & {\rm for} & n,m=0 
\eaa
\right.\non\\
&&\int_0^{2\pi R}\sin(m_ny)\cos(m_my)dy=0\non\\
&&\int_0^{2\pi R}\cos(m_ny)\cos(m_my)
\cos(m_ly) dy=\frac{\pi R}{2}  A_{nml}\non\\
&&\int_0^{2\pi R}\sin(m_ny)\sin(m_my)
\cos(m_ly) dy=\frac{\pi R}{2}  C_{nml}\non\\
&&\int_0^{2\pi R}\cos(m_ny)\cos(m_my)
\sin(m_ly) dy=0\non\\
&&\int_0^{2\pi R}\cos(m_ny)\cos(m_my)
\cos(m_ky) \cos(m_ly) dy=\frac{\pi R}{4} B_{nmkl}\non\\
&&\int_0^{2\pi R}\sin(m_ny)\sin(m_my)
\cos(m_ky) \cos(m_ly) dy=\frac{\pi R}{4} D_{nmkl}
\eea
where
\bea
A_{nml} &\equiv& \delta_{l,n+m}+\delta_{l,n-m}+\delta_{l,-n+m}+
\delta_{l,-n-m}\non\\
C_{nml} &\equiv& -\delta_{l,n+m}+\delta_{l,n-m}+
\delta_{l,-n+m}-\delta_{l,-n-m}\non\\
B_{nmkl} &\equiv& \delta_{l,-n-m+k}+\delta_{l,n+m-k}+ \delta_{l,-n+m+k} +\delta_{l,n-m-k}
\non\\
&&+\delta_{l,n-m+k}+\delta_{l,-n+m-k}
+\delta_{l,n+m+k} +\delta_{l,-n-m-k} \non\\
D_{nmkl} &\equiv& \delta_{l,-n+m-k}+\delta_{l,-n+m+k}+ \delta_{l,n-m+k} +\delta_{l,n-m-k}
\non\\
&&-\delta_{l,n+m-k}-\delta_{l,n+m+k}
-\delta_{l,-n-m+k} -\delta_{l,-n-m-k}
\eea

\renewcommand{\theequation}{B.\arabic{equation}}
\setcounter{equation}{0}
\noindent \hspace*{-0.72cm}
\centerline{\bf APPENDIX B}

Since in the case of mixing between KK modes the standard technique developed 
in ref.\cite{Delgado:1998qr} for a calculation of the effective potential can not be applied directly, we
present here some details of the derivation that lead to the result shown in eq.(\ref{eff_mix}). 
In a case of non-diagonal mass matrix $M^2$ we have to consider the following form of 
the effective potential in  Euclidean space:
\beq
V(\ph)=\frac12 Tr\left\{ \int\frac{d^4p}{(2\pi)^4}\sum_{n=-\infty}^{\infty}\ln\left[l^2(p^2+
M^2)\right]\right\}\,,
\eeq
where $M$ is in general non-diagonal mass matrix for KK modes and we have restricted 
ourself to the no-shift case: $\omega=0$. For the $(A_{5\,n},\chi_n)$ system we have
\beq
M^2=\left(
\baa{cc}
e^2\ph^2 & -e\ph m_n\\
-e\ph m_n & \mu^2+\lambda \ph^2 + m_n^2
\eaa
\right)\,.
\eeq
Going to diagonal form of $M^2$ it is easy to see that
\beq
{\rm Tr}\left\{\ln[l^2(p^2+M^2)]\right\}=\ln[l^4(p^4+p^2{\rm Tr} M^2+{\rm Det} M^2)]\,.
\eeq
Since 
\beq
{\rm Tr} M^2= e^2\ph^2 +\mu^2+\lambda \ph^2 + m_n^2 \lsp {\rm and} \lsp 
{\rm Det} M^2=e^2\ph^2(\mu^2+\lambda \ph^2 )
\eeq
we obtain eventually
\beq
{\rm Tr}\left\{\ln[l^2(p^2+M^2)]\right\}=\ln\left[l^2E^2+n^2\pi^2\right]\,,
\eeq
where irrelevant constant terms have been dropped and
\beq
E^2=p^2+a+\frac{b}{p^2}\,,
\eeq
with
\beq
a=e^2\ph^2 +\mu^2+\lambda \ph^2  \lsp {\rm and} \lsp
b=e^2\ph^2(\mu^2+\lambda \ph^2 )\label{ab}
\eeq
Following the method adopted for diagonal mass matrices, one needs to differentiate 
$W\equiv \frac12\sum_{n=-\infty}^{\infty}\ln\left[(lE)^2+n^2\pi^2\right]$
with  respect to $E$, then trade the summation
for a contour integral and eventually integrate over $E$. The result is 
\beq
W=lE+\ln\left(1-e^{-2lE}\right) + {\rm constant}
\label{w}
\eeq
The term that is ultraviolet divergent for a cut-off regularization emerges from the
integral of the first term in eq.(\ref{w}):
\beq
V^{(\infty)}_{mix}=l\int\frac{d^4p}{(2\pi)^4}\sqrt{p^2+a+\frac{b}{p^2}}
\label{mix_infty}
\eeq
The compactification radius dependent contribution consists of the integral of
the second term in eq.(\ref{w}):
\beq
V^{(R)}_{mix}=\int\frac{d^4p}{(2\pi)^4}\ln\left(1-e^{-2lE}\right)
\label{mix_r}
\eeq
The following formula will be adopted
\beq
\int_0^\infty\frac{x^{\alpha-1}dx}{(ax^2+2bx+c)^\rho}=a^{-\frac{\alpha}{2}}c^{\frac{\alpha}{2}-\rho}
{\rm B}(\alpha,2\rho-\alpha){\rm F}\left(\frac{\alpha}{2},\rho-\frac{\alpha}{2};\rho+\frac12;1-\frac{b^2}{ac}\right)\,,
\label{integ}
\eeq
where ${\rm B}(x,y)$ and ${\rm F}(a,b;c;z)$ are the Euler beta function and hypergeometric 
function, respectively. Using the above result one can show that for the dimensional regularization
the integral in eq.(\ref{mix_infty}) is finite in the limit $n\to 4$ and the corresponding potential 
reads\footnote{It could be verified that the following result reproduce the formula
(\ref{vinftyres}) in the limit $b\to 0$.}:
\beq
V^{(\infty)}_{mix}=-\frac{y^{1/2}(y^2-1)x^5}{2^{12}\sqrt{2}\pi^5R^4}
{\rm F}\left(-\frac14,\frac74;2;1-\frac{1}{y^2}\right)\,,
\label{mix_infty_veff}
\eeq
where $x$ and $y$ are defined  in the main text, see eq.(\ref{xy}).

The integral $V^{(R)}_{mix}=\int\frac{d^4p}{(2\pi)^4}\ln\left(1-e^{-2lE}\right)$
is more difficult to perform, so we will adopt an asymptotic expansion in the limit
$2\pi R\ph \to \infty$ that is an excellent approximation in the region of our 
interest\footnote{Since we are interested in the stability of the vacuum, therefore
it is enough to know the shape of the effective potential for $\ph \sim {\rm few} \tev$,
what turns out to be sufficient for the application of the asymptotic expansion of the 
integral.}. The result reads
\beq
V^{(R)}_{mix}\simeq -\frac{y^{3/2}(1+y)^{1/4}x^{7/2}}{2^9\pi^5\sqrt{\pi}R^4}
{\rm Li}_{\frac32}\left(e^{-x\sqrt{1+y}}\right)\,.
\label{mix_r_veff}
\eeq
Eventually, the contribution to the effective potential from the $(A_{5\,n},\chi_n)$ system 
is the following:
\beq
V_{eff}^{(A_5,\chi)}=\frac12\left(V^{(\infty)}_{mix}+V^{(R)}_{mix}-V_{0\, finite}^{(A_0)}
-V_{0\, finite}^{(\chi_0)}\right)\,,
\label{vmixtot}
\eeq
where 
$V_{0\, finite}^{(A_0)}$ and $V_{0\, finite}^{(\chi_0)}$
are the finite parts of scalar contributions (see eq.(\ref{veffmsbar})) to the effective potential calculated for 
the zero mode vector boson mass ($m_{A_0}^2=e^2\ph^2$)
and Goldstone boson ($m_{\chi_0}^2=\mu^2+\lambda \ph^2$), respectively.

\end{document}